\documentclass[12pt]{iopart}
\usepackage{graphicx}
\usepackage{iopams}
\usepackage[pdfborder=000, colorlinks, linkcolor=blue, citecolor=blue]{hyperref}
\usepackage[numbers,sort&compress]{natbib}
\usepackage{color}
\usepackage{caption}
\usepackage{algorithmicx}
\usepackage{algorithm}
\usepackage{float}
\usepackage{algpseudocode}
\usepackage{fancyhdr}
\begin{document}

\title[]{Probabilistic activity driven model of temporal simplicial networks and its application on higher-order dynamics}

\author{Zhihao Han$^{1,3}$, Longzhao Liu$^{2,3,4,5,6,7,*}$, Xin Wang$^{2,3,4,5,6,7,*}$, Yajing Hao$^{1,3}$, Hongwei Zheng$^{6,10}$, Shaoting Tang$^{2,3,4,5,6,7,8,9}$, Zhiming Zheng$^{2,3,4,5,6,7,8,9}$}

\address{$^1$ School of Mathematical Sciences, Beihang University, Beijing 100191, China}
\address{$^2$ Institute of Artificial Intelligence, Beihang University, Beijing 100191, China}
\address{$^3$ Key laboratory of Mathematics, Informatics and Behavioral Semantics (LMIB), Beihang University, Beijing 100191, China}
\address{$^4$ State Key Lab of Software Development Environment (NLSDE), Beihang University, Beijing 100191, China}
\address{$^5$ Zhongguancun Laboratory, Beijing, P.R.China}
\address{$^6$ Beijing Advanced Innovation Center for Future Blockchain and Privacy Computing, Beihang University, Beijing 100191, China}
\address{$^7$ PengCheng Laboratory, Shenzhen 518055, China}
\address{$^8$ Institute of Medical Artificial Intelligence, Binzhou Medical University, Yantai 264003, China}
\address{$^9$ School of Mathematical Sciences, Dalian University of Technology, Dalian 116024, China}
\address{$^{10}$ Beijing Academy of Blockchain and Edge Computing (BABEC), Beijing 100085, China}

\ead{longzhao@buaa.edu.cn, wangxin\_1993@buaa.edu.cn}

\vspace{10pt}
\begin{indented}
\item[]
\end{indented}

\begin{abstract}
Network modeling characterizes the underlying principles of structural properties and is of vital significance for simulating dynamical processes in real world. However, bridging structure and dynamics is always challenging due to the multiple complexities in real systems. Here, through introducing the individual's activity rate and the possibility of group interaction, we propose a probabilistic activity driven (PAD) model that could generate temporal higher-order networks with both power-law and high-clustering characteristics, which successfully links the two most critical structural features and a basic dynamical pattern in extensive complex systems. Surprisingly, the power-law exponents and the clustering coefficients of the aggregated PAD network could be tuned in a wide range by altering a set of model parameters. We further provide an approximation algorithm to select the proper parameters that can generate networks with given structural properties, the effectiveness of which is verified by fitting various real-world networks. Lastly, we explore the co-evolution of PAD model and higher-order contagion dynamics, and analytically derive the critical conditions for phase transition and bistable phenomenon. Our  model provides a basic tool to reproduce complex structural properties and to study the widespread higher-order dynamics, which has great potential for applications across fields.
\end{abstract}

%
%
%
%
%

\section{Introduction}

Network structure not only directly reflects interaction patterns in complex systems, but also significantly affects the dynamical outcomes of these systems such as opinion formation~\cite{li2006dynamics,zhan2019impact,wu2004social,wang2020public,liu2020homogeneity}, disease spreading~\cite{holme2016temporal,keeling2005implications,hazarie2021interplay}, brain dynamics~\cite{lynn2019physics,schmalzle2017brain}, ecosystem evolution~\cite{guimaraes2020structure,wang2020eco} and adoption of innovation~\cite{cowan2004network,iacopini2018network}. Therefore, generating networks with specified or real structure, i.e. network modeling, is of vital importance in exploring and controlling networked systems. In this field, there are two widely-used classes of network models: connectivity-driven model and activity-driven model. The connectivity-driven model takes the structural patterns of networks as the basis for the formation mechanism, whose typical cases include Erd\"{o}s-R\'{e}nyi Model~\cite{erdHos1960evolution}, Watts-Strogatz Model~\cite{watts1998collective}, and Barab\'{a}si-Albert Model~\cite{barabasi1999emergence}. The activity-driven model encodes the network structure into nodes' property, i.e., the activity rate, and generate time-varying networks which provide a natural way to study the co-evolution of network structure and dynamical processes~\cite{perra2012activity,pozzana2017epidemic}. Though both kinds of models can reflect part of real topological features like power-law degree distributions~\cite{starnini2013topological}, they simply consider pairwise interactions as the only underlying block of generating complex networks.

In recent years, higher-order interactions, usually represented by hyperedges and simplexes, have been proved to exist widely in real systems~\cite{grilli2017higher,battiston2020networks}. In particular, compared to traditional pairwise networks, the dynamics, including spreading dynamics~\cite{liu2021multilayer,ferraz2021phase,de2020social}, evolutionary dynamics~\cite{alvarez2021evolutionary,civilini2021evolutionary} and synchronization~\cite{skardal2020higher,gambuzza2021stability,ghorbanchian2021higher}, on the top of higher-order networks all display fundamentally different results~\cite{majhi2022dynamics}. For example, Iacopini $et$ $al$ incorporated the transmission mechanism occurring in simplexes, and found the emergence of bistable phenomena in social contagion processes, while this phenomena can not take place in pairwise networks~\cite{iacopini2019simplicial}. St-Onge $et$ $al$ considered the higher-order structure of contacts in epidemic spreading processes and found the pattern of superexponential spread~\cite{st2022influential}. It can be concluded that higher-order interactions play a pivotal role in both structure and dynamics of networked systems. Accordingly, a basic and prominent problem is how to generate higher-order networks with desired structures that could support the important higher-order interactions.

Several effective higher-order network models have been proposed to stress this issue~\cite{bianconi2016network,courtney2018dense,barthelemy2022class,costa2016random}. For example, Petri $et\ al$ proposed simplicial activity driven (SAD) model by taking simplexes as the underlying blocks of higher-order~\cite{petri2018simplicial}. T. Courtney $et\ al$ presented a nonequilibrium model for weighted simplicial complexes with nontrivial topology (manifolds and heterogeneous scale-free degree distribution), and the generalized strength can grow linearly, superlinearly, or exponentially~\cite{courtney2017weighted}. Kovalenko $et\ al$ combined preferential and nonpreferential attachment mechanisms to grow simplicial complexes, which are characterized by scale-free degree distribution and an either bounded or scale-free generalized degree distribution~\cite{kovalenko2021growing}. Despite the progress, there is still a long way from generating higher-order networks with specified or real structures. Note that real-world networks, such as social networks, usually have power-law and high-clustering characteristics, and that the power-law exponents of different networks have a wide value range (usually 2 - 3)~\cite{albert2002statistical}. Connectivity-driven model and activity-driven model can not reflect these two characteristics simultaneously and do not support higher-order dynamics. The existing higher-order network models, such as SAD model, display high-clustering characteristics but can not generate a wide range of power-law exponents. In short, it remains unknown how to generate networks which not only support the co-evolution with the ubiquitous higher-order dynamics but also have arbitrary power-law exponents and clustering coefficients in order that a large number of real-world networks could be reconstructed.

To fill this gap, we propose a probabilistic activity driven (PAD) model by considering individual's activity rate and the possibility of group interactions, usually being represented as simplexes, in the processes of network evolution. The aggregated networks generated by the PAD model can have both power-law and high-clustering characteristics. More importantly, model parameters can be utilized to alter the aggregated networks' power-law exponents and clustering coefficients for a wide range. Furthermore, we provide an approximation algorithm to select model parameters that can generate networks with specific structural properties and verify its effectiveness by reconstructing three real-world networks with different topological properties. Finally, we show how complex dynamical behaviors emerge from the  coupling dynamics of higher-order contagion and network evolution based on PAD model. In particular, we identify stable and unstable equilibrium manifolds and theoretically derive the corresponding thresholds of phase transitions.

\section{Probabilistic activity driven model}

Higher-order interactions have been proved as indispensable patterns in networked systems, such as group interactions in social networks~\cite{cencetti2021temporal} and cooperations of authors in scientific collaboration networks~\cite{vasilyeva2021multilayer}. In network sciences, the higher-order interactions are usually represented by simplexes. Specifically, a $k$-simplex $\sigma$ means an unordered set of $k+1$ vertices $\{v_0,v_1,\dots,v_k\}$ where any vertices connect each other~\cite{salnikov2018simplicial}. For example, a 2-simplex is a 'full-triangle' which includes not only the hyperedges among all nodes but also the pairwise links between any vertices. This cognition naturally motivates us to adopt simplexes as the underlying blocks in network modeling.

In addition, complying with previous studies, we also consider activity rate of vertices, denoted by $a$, to measure their dynamical property, which is a widely-existing feature in real systems. For example, this parameter can represent not only the individuals' willingness in social networks but also authors' efficiency in collaboration networks.

Overall, we propose probabilistic activity driven model (PAD model) by considering the activity rate of vertices and the possibility of simplexes as the fundamental blocks to generate temporal higher-order networks. In this model, each node $i$ is assigned two properties: activity rate $a_i$($\epsilon\le a_i\le 1$) and possibility of interactions on simplexes $p_i$($0\le p_i\le 1$). Here, activity rate $a_i$ represents the possibility that node $i$ is activated and create new contacts per unit time. $p_i$ denotes the probability that active nodes participate higher-order interactions, i.e., creating simplexes rather than pairwise links. Suppose that the two parameters obey joint probability distribution $H(a,p)$ . Then the network model can be described as the following steps (see figure \ref{sketch}) :

\begin{figure}
	\center
	\includegraphics[width=12.5cm]{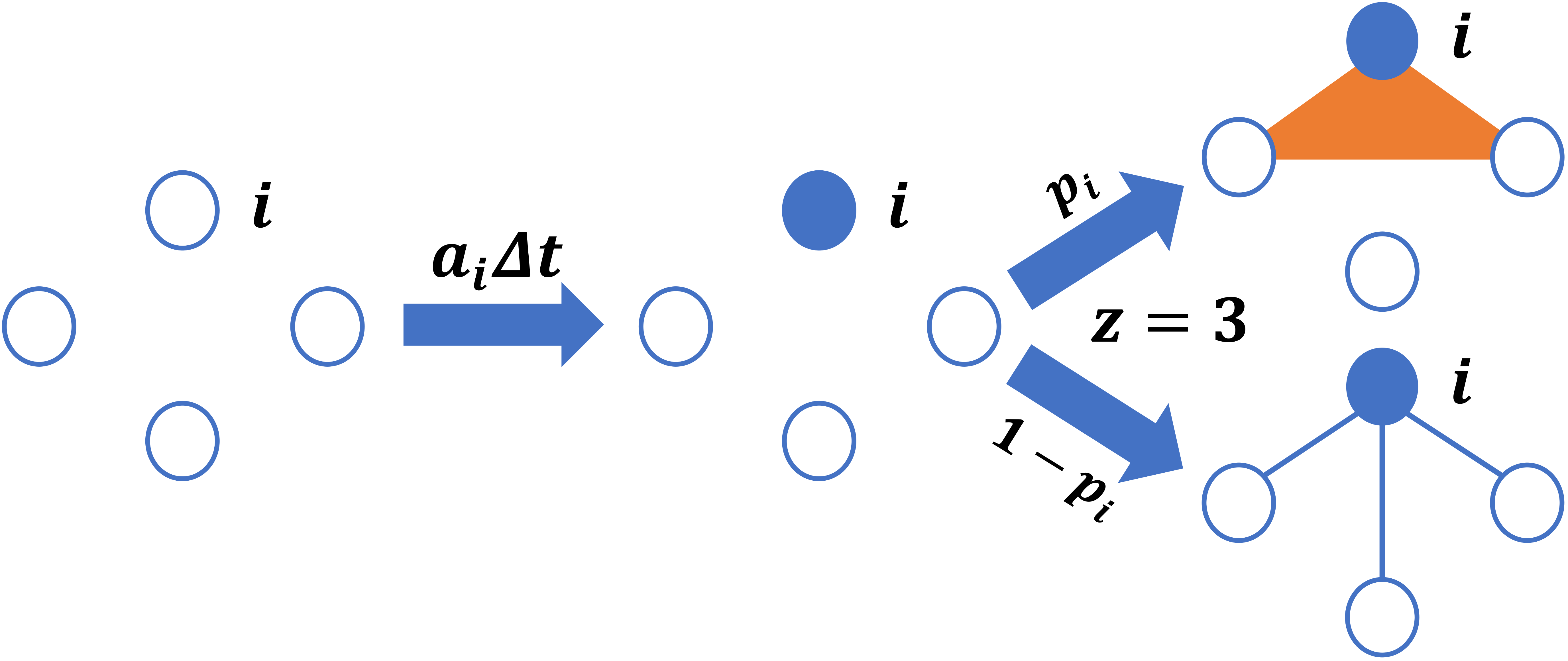}
	
	\caption{PAD model. At each time step, a node $i$ activates with probability $a_i\Delta t$.  Upon activation it creates a ($z-1$)-simplex with probability $p_i$; with probability $1-p_i$, it generates $z(z-1)/2$ 1-simplex (link).}
	\label{sketch}
\end{figure}

\begin{enumerate}
	\item[(i)] At each time step $t$, the instantaneous network $G_t$ starts with $N$ disconnected nodes.
	\item[(ii)]Each node $i$ activates with probability $a_i\Delta t$. When node $i$ is activated, it has two options of creating new connections: with probability $p_i$, it creates a $(z-1)$-simplex with $(z-1)$ other nodes chosen randomly; with probability $1-p_i$, it generates $m$ links (1-simplex) that are connected to $m$ other randomly selected nodes. In order to ensure the consistency in the number of interactions, we make $m=z(z-1)/2$. Here, the size $z$ follows a discrete distribution $p(z)$. The distribution of $m$ can be obtained from the relation between $z$ and $m$.
	\item[(iii)]At the next time step $t+\Delta t$, the existing network structures are erased and the process starts anew.
\end{enumerate}

For simplicity, we call the higher-order networks generated by probabilistic activity driven model as PAD networks.

\section{Structural properties of aggregated PAD networks}\label{network structure}

Here, we define the aggregated PAD networks as the union of all instantaneous network $G_t$ generated by PAD model in $T$ time steps~\cite{perra2012activity}, i.e., denoted by $G_T=\bigcup^{t=T}_{t=0}G_t$. For simplicity, we set a time step $\Delta t$ as 1. In this section, through mean-field theory and simulations, we mainly explore the structural properties of aggregated PAD networks, such as degree distribution, clustering coefficient and higher-order structure.

We define $k_n(i,T)$ as the number of distinct $n$-simplex which $i$ belongs to in aggregated PAD network $G_T=\bigcup^{t=T}_{t=0}G_t$. Then, $k_1(i,T)$ represents the degree of node $i$, i.e, the number of distinct nodes that interact with node $i$ at least once during $T$. By utilizing mean-field theory (see \ref{distribution}, for details), $k_1(i,T)$ can be approximated by

\begin{equation}
\eqalign{
		k_1(i,T)&\simeq N(1-e^{-\frac{T(a_id_i+\langle a\rangle\langle d\rangle+h)}{N}})
	}
\label{degree equation}
\end{equation}

where $d_i=\langle z-1\rangle p_i+(1-p_i)\langle m\rangle,\langle d\rangle=\langle z-1\rangle\langle p\rangle+\langle 1-p\rangle\langle m\rangle,h=\langle (z-1)(z-2)\rangle\langle ap\rangle$.

Assume that all nodes have homogeneous possibility of creating $(z-1)$-simplex, i.e, $p_i=p$ for all node $i$. Under this condition, given the arbitrary distribution $F(a)$ of activity rate, we can derive the degree distribution of the aggregated PAD network, which read

\begin{equation}
\eqalign{
	P_T(k)\sim\frac{1}{T\langle d\rangle(1-\frac{k}{N})}F\left[-\frac{N}{T\langle d\rangle}\ln(1-\frac{k}{N})-\langle a\rangle-\frac{h}{\langle d\rangle}\right]
}
\label{ori 1-order degree distribution}
\end{equation}

Considering that $N\gg k$ in real-world networks, equation (\ref{ori 1-order degree distribution}) can be approximated by

\begin{equation}
\eqalign{
	P_T(k)\sim\frac{1}{T\langle d\rangle}F\left[\frac{k}{T\langle d\rangle}-\langle a\rangle-\frac{h}{\langle d\rangle}\right]
}
\label{eq3}
\end{equation}

Equation (\ref{eq3}) shows that the degree distribution of aggregated networks has similar features with the distribution of activity rate. In other words, our model can approximately generate aggregated networks following power-law rule by setting that activity rate obeys power-law distribution.

Then we study the higher-order structural properties of aggregated networks, i.e., the expected number $k_n(i,T)$ of $n$-simplex to which a node $i$ belongs. $k_n(i,T)$ can be calculated in a similar way of deriving $k_1(i,T)$. Here we present the computational formula of $k_2(i,T)$, which is as follows:

\begin{equation}
\eqalign{
		k_2(i,T) \simeq {N-1 \choose 2}\left[1-e^{-{\frac{\langle (z-1)(z-2)\rangle a_ip_iT+\langle (z-1)^2(z-2)\rangle\langle ap\rangle T}{(N-1)(N-2)}}}\right]
}
	\label{simplex equation}
\end{equation}

\begin{figure}
	\center
	\includegraphics[width=12.5cm]{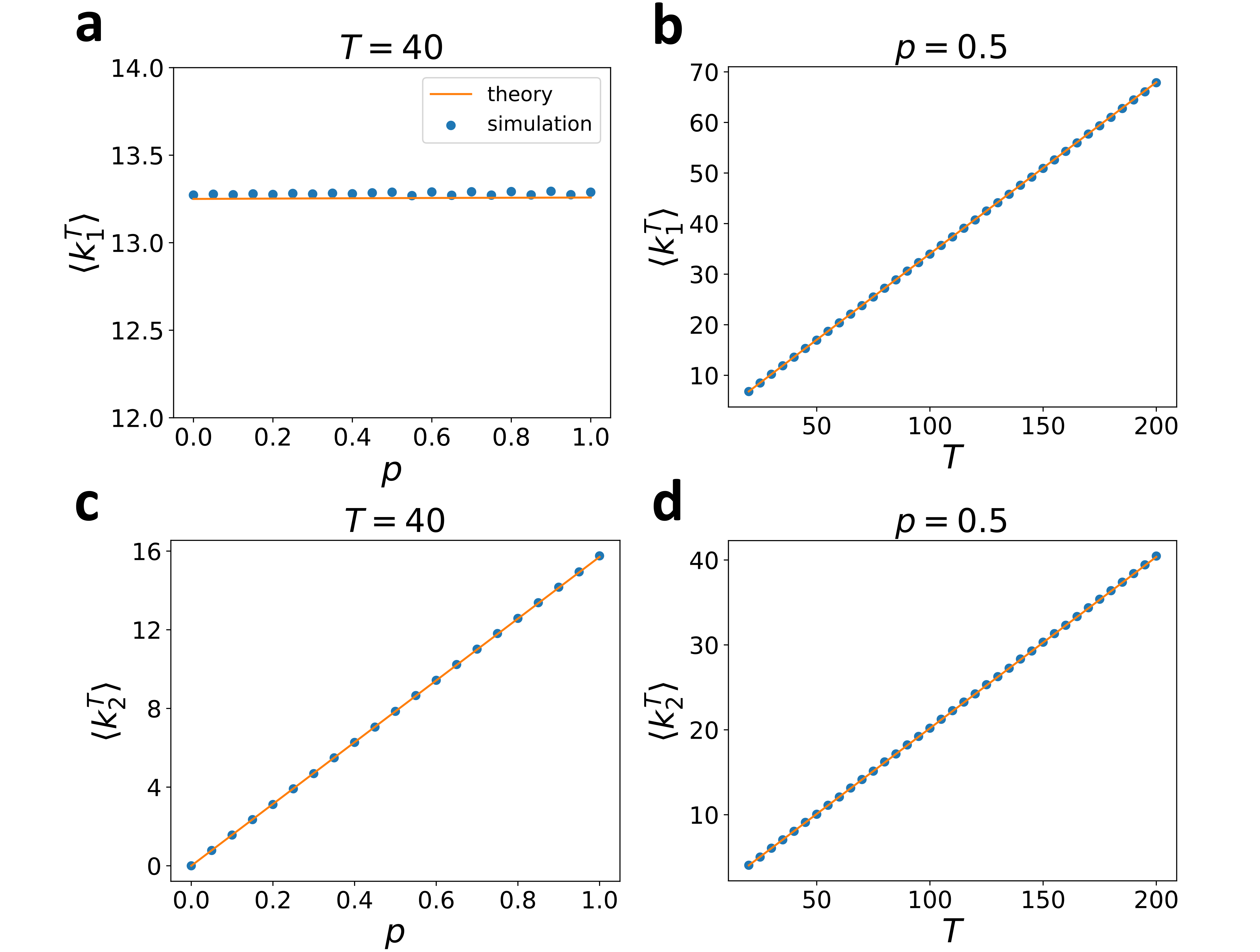}
	
	\caption{Degree $\langle k_1^T\rangle$ and 2-simplex degree $\langle k_2^T\rangle$ of aggregated networks. We present $\langle k_1^T\rangle$ and 2-simplex $\langle k_2^T\rangle$ of aggregated networks with respect to different $p$ and $T$. Dots correspond to results averaged by 10 independent simulations, while solid lines are theoretical predictions solved by equation (\ref{degree equation}) ((a), (b)) and equation (\ref{simplex equation}) ((c), (d)). Parameters: $N=30000$, $F(a)\propto a^{-1.3}$; (a)(c) $T=40$; (b)(d) $p=0.5$.}
	\label{degree}
	
\end{figure}

\begin{figure}
	\center
	\includegraphics[width=12.5cm]{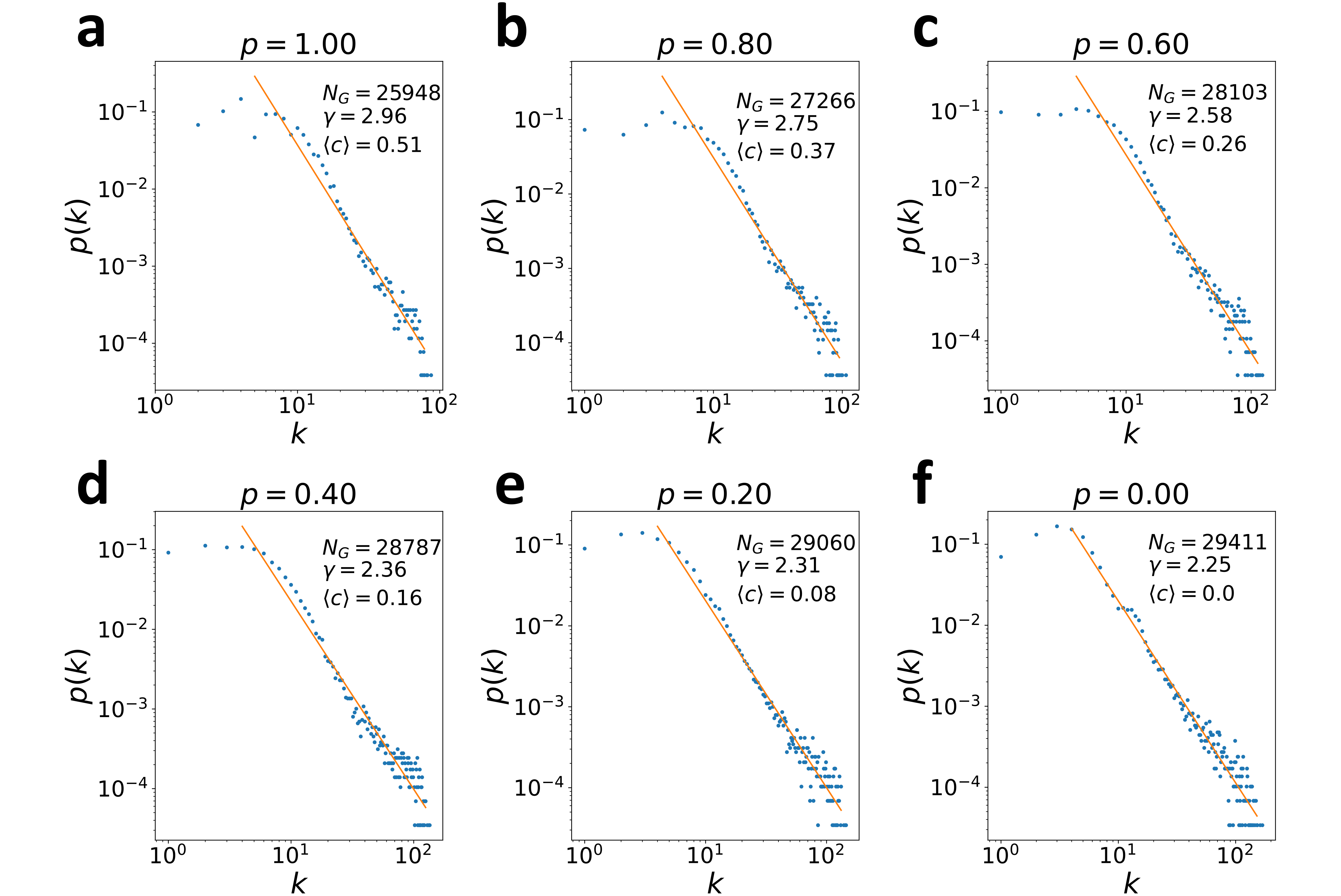}
	
	\caption{Degree distribution of the aggregated PAD networks. We set the model parameter $p$ as different values and respectively generate corresponding PAD networks, whose degree distributions are presented in the plots. Results show that all networks follow power-law rule. The power-law exponents and clustering coefficients are represented by $\gamma$ and $\langle c\rangle$, respectively. $N_G$ represents the size of largest connected component of networks. Other parameters: $N=30000$, $T=22$, $p(z=l)=1/3(l=3,4,5)$ and $F(a)\propto a^{-1.7}$.}
	\label{degree distribution over p}
\end{figure}

Here, we first verify the effectiveness of our theoretical analysis and present the structure of aggregated networks  by simulating the PAD model. Specifically, figure \ref{degree}(a) and figure \ref{degree}(b) show the average degree $\langle k_1^T\rangle$ as a function of probability $p$ and aggregation steps $T$, respectively. The average degree depends on $T$ but is not affected by $p$. This indicates that our model settings successfully ensure the invariance of average degree no matter what $p$ is. Figure \ref{degree}(c) and figure \ref{degree}(d) show that the 2-simplex degree $\langle k_2^T\rangle$ increases with $p$ and $T$ growing. Noteworthy, parameter $p$ provides a way of altering $\langle k_2^T\rangle$ without changing $\langle k_1^T\rangle$. Moreover, figure \ref{degree}(a)-(d) all show that the theoretical predictions are in agreement with simulation results, which proves the effectiveness of our theoretical calculations.

\begin{figure}
	\center
	\includegraphics[width=12.5cm]{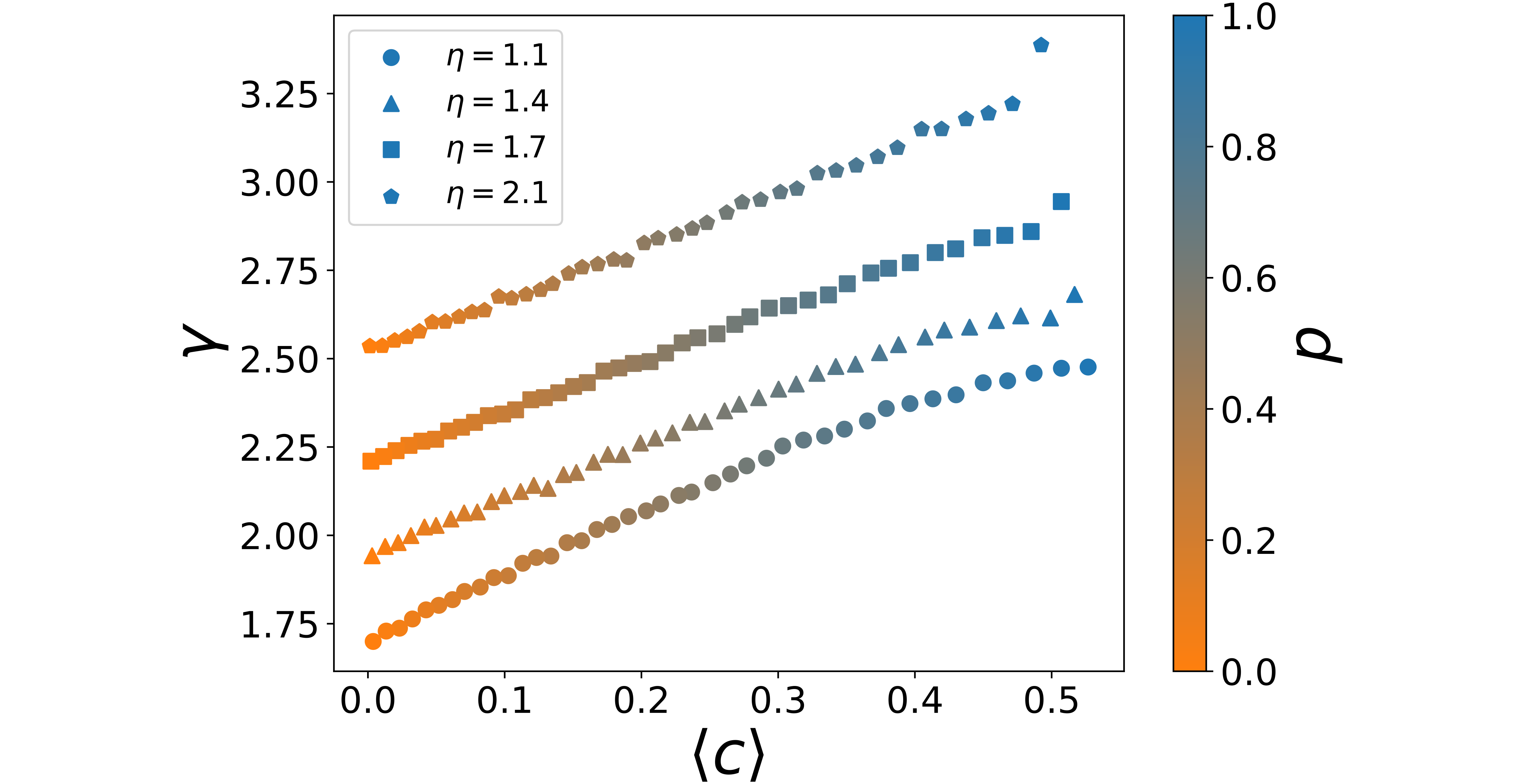}
	
	\caption{Phase diagram about structural properties of aggregated PAD networks. Structural properties of networks, like clustering coefficients and power-law exponents, are presented under different model parameters, i.e., possibility of creating higher-order simplexes $p$ and power-law exponent of activity rate $\eta$. Under each combination of model parameters, we can generate a network, represented by a point in the plot, and it is ensured that the size of GCC is more than $85\%$ of the network. Parameters: $N=30000$, $T=22$, and $F(a)\propto a^{-\eta}$.}
	\label{phase structure p k}
\end{figure}

To give a more intuitive presentations, we simulate our network model to generate PAD networks with parameter $p$ varying and $F(a)\propto a^{-1.7}$. Here, we use the properties of largest connected component (GCC) to represent network properties. Figure \ref{degree distribution over p} presents the degree distribution of largest connected component (GCC) of these aggregated PAD networks. Noteworthy, all networks follow power-law degree distribution no matter what $p$ is, which is consistent with the cognition from equation (\ref{eq3}).  Moreover, the average clustering coefficients $\langle c\rangle$ of these networks can increase to 0.5 with $p$ growing from 0 to 1. These indicate that our network model can generate networks with both power-law degree distribution and high-clustering characteristics, which are widely-existing structural features of real-world networks.

Furthermore, given that $F(a)\propto a^{-\eta}$, we comprehensively explore the properties of aggregated PAD networks under different combinations of model parameters: possibility of creating higher-order simplexes $p$ and power-law exponent of activity rate $\eta$. Under each combination of parameters, we use PAD model to generate a network, whose power-law exponent ($\gamma$, vertical axis) and clustering coefficients ($\langle c\rangle$, horizontal axis) are presented as a point in figure \ref{phase structure p k}. Results show that both $\gamma$ and $\langle c\rangle$ monotonously increase with $p$ and $\eta$ growing. Surprisingly, the generated networks have a wide range of power-law degree exponents $\gamma$ and clustering coefficients $\langle c\rangle$, i.e., $(\gamma, \langle c\rangle)\in [2.3,3.3]\times [0,0.5]$. That is, for any given network property $(\gamma_0, c_0)$ in the above interval, our model can generate corresponding synthetic network through setting appropriate $p$ and $\eta$. It indicates the huge potentials of PAD model to generate networks with specified structure. Naturally, this understanding also leads to the following question: how to select suitable model parameters to generate specified networks.

\begin{figure}
	\centering
	\includegraphics[width=12.5cm]{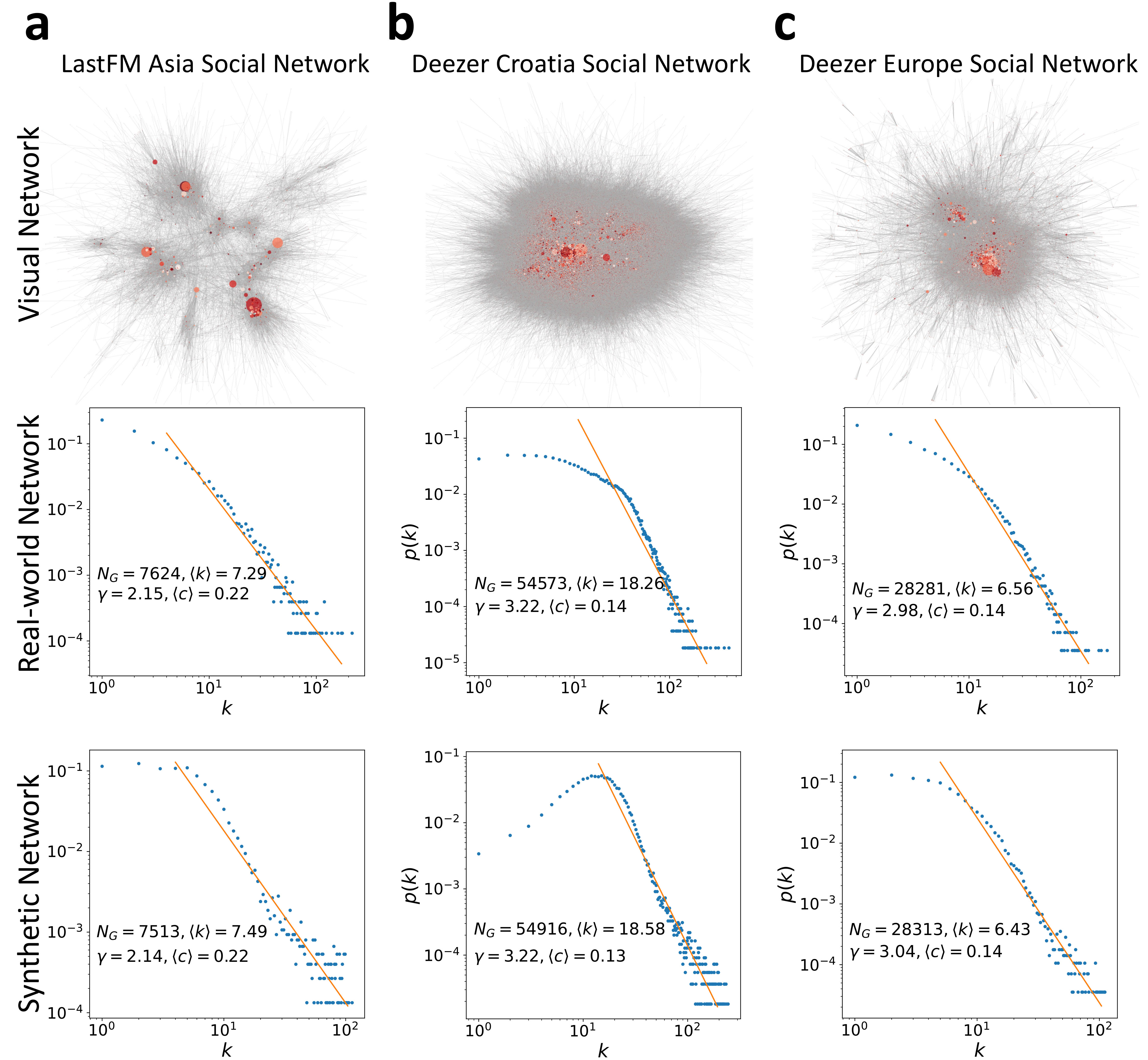}
	\caption{Fitting the structural properties of real-world networks. We consider three real-networks. Each column respectively represent real network structural properties and the fitting results generated by PAD model. The structural properties (network size $N_G$, average degree $\langle k \rangle$, degree distribution $\gamma$ and average clustering coefficient $\langle c \rangle$) are annotated in each subgraph.}
	\label{fit network}
\end{figure}

Thus, we also provide an algorithm of selecting suitable model parameters, such as activity rate distribution, possibility of creating higher-order simplex and aggregation steps, to ensure that the generated networks have specified or real structural properties. Its pseudo-code can be found in \ref{fitting method}. To verify the effectiveness of this algorithm, we try to use this algorithm and PAD model to fit the structural properties of three different real-world network datasets~\cite{snapnets} (see \ref{data descripution} for details ): Deezer Europe Social Network, Deezer Croatia Social Network and LastFM Asia Social Network, which have significantly different power-law exponents ranging from 2 to 3. Figure \ref{fit network} presents the structural properties extracted from the real networks and the fitted networks. All fitting results show that the structural properties of synthetic networks, such as degree distribution and clustering coefficient, are almost the same with the real networks. To some extent, this proves that for a wide range of structural properties, the combination of our model and this algorithm can be used to approximately generate corresponding networks.

\section{Higher-order social contagion on temporal simplicial networks}\label{sec3}

In addition to generating simplicial network with specified power-law exponent and clustering coefficient, our model also has the advantage of coupling with higher-order dynamics at the same time scale. Here we take the co-evolution of higher-order contagion dynamics and our model as typical case of analysis. Specifically, we use the higher-order ignorant-spreader-ignorant (SIS) contagion model~\cite{iacopini2019simplicial}. In this model, individuals are divided into 2 classes: spreader($S$) and ignorant ($I$).  Spreader ($S$) represents individuals who adopt norms and are willing to spread, and ignorant ($I$) stands for individuals who do not adopt norms or have no motivation to spread. There are two main dynamical mechanisms, i.e., recovery and transmission. (1) (recovery) spreader becomes ignorant with probability $S\stackrel{\mu}{\longrightarrow}I$. (2) (transmission) For any $R$-simplex, its node $i$ in state $I$ changes the state to spreader with probability $\beta_R$ if all other $R$ nodes in this simplex are spreaders, i.e., $Simplex(I,RS)\stackrel{\beta_R}{\longrightarrow}Simplex((R+1)S)$. In this section, we will detailedly explore the coupling dynamics between higher-order SIS model and our PAD model.

\subsection{Theoretical framework}

Here, we explore dynamical equations of higher-order social contagion on temporal simplicial networks generated by PAD model. Consider a PAD network $G$ with $N$ nodes. We define $N_{a,p}^t$ as the number of nodes whose activity rate and possibility of creating higher-order simplex are $a$ and $p$ at time $t$. Thus, the density of nodes $n_{a,p}^t$ is equal to $N_{a,p}^t/N$. Let $i_{a,p}^t$ and $s_{a,p}^t$ represent the density of ignorants and spreaders, respectively. Clearly, $i_{a,p}^t+s_{a,p}^t=n_{a,p}^t$. By utilizing mean-field theory, the dynamical equations of $R$-order contagion describing how the density of spreaders evolves with time varying can be written as
\begin{equation}
	\eqalign{
		&s_{a,p}^{t+\Delta t}-s_{a,p}^t = -\mu \Delta ts_{a,p}^t + \beta_1\Delta ti_{a,p}^ta(1-p)\int P(m)m dm\int s_{a^{'},p^{'}}^tda^{'}dp^{'} \\
		&+\beta_1\Delta ti_{a,p}^t\int P(m)m dm\int a^{'}(1-p^{'}) s_{a^{'},p^{'}}^tda^{'}dp^{'}\\
		&+\sum_{w=1}^{R} \beta_w\Delta t i_{a,p}^tap\int P(z){z-1 \choose w}dz(\int s_{a^{'},p^{'}}^tda^{'}dp^{'})^w\\
		&+\sum_{w=1}^{R} \beta_w\Delta t i_{a,p}^t\int a^{'}p^{'}s_{a^{'},p^{'}}^tda^{'}dp^{'}\int P(z)(z-1){z-2 \choose w-1}dz(\int s_{a^{''},p^{''}}^tda^{''}dp^{''})^{w-1}\\
		&+\sum_{w=1}^{R} \beta_w\Delta ti_{a,p}^t\int a^{'}p^{'}n_{a^{'},p^{'}}^tda^{'}dp^{'}\int P(z)(z-1){z-2\choose w}dz(\int s_{a^{''},p^{''}}^tda^{''}dp^{''})^{w},
}
\label{social contagion equation}
\end{equation}

The first term corresponds to the recovery process of the spreader. The second term represents the ratio that the activated ignorant nodes actively generate links connecting to spreaders and are infected by these spreaders. The third term means the ratio that inactivated ignorant nodes are passively receive the connections from activated spreaders and are infected. These two terms correspond to the transmission process through 1-simplex (i.e., $1$-order contagion). The last three terms correspond to the transmission process through $w$-simplex for $1\le w\le R$. Similarly, the forth term is the ratio that activated ignorant nodes actively create a $(z-1)$-simplex including $w$ spreaders and are infected. The fifth and sixth term stems from the fact that ignorant is passively connected by activated spreaders and activated ignorant nodes creating a $(z-1)$-simplex, respectively, and is infected.

Next, we take $2$-order contagion on PAD networks as typical cases to analyze the evolutionary results of this coupling dynamics.

\subsubsection{Homogeneous situation}

To begin with, we consider homogeneous situations where all nodes have the homogeneous properties including the same activity rate $a$ and the same possibility of creating higher-order simplexes. Under this homogeneous condition, we can directly write the dynamical equation of $2$-order contagion on homogeneous PAD networks according to equation (\ref{social contagion equation}), which is
\begin{equation}
	\eqalign{
		\frac{d\rho}{dt} &= -\mu\rho+2\beta_1a(1-p)\langle m\rangle\rho(1-\rho)+2\beta_1ap\langle z-1\rangle\rho(1-\rho)\\
		&+\beta_1ap\langle(z-1)(z-2)\rangle\rho(1-\rho)+\frac{3\beta_2ap\langle(z-1)(z-2)\rangle\rho^2(1-\rho)}{2}\\
		&+\frac{\beta_2ap\langle(z-1)(z-2)(z-3)\rangle\rho^2(1-\rho)}{2}
	}
\label{2-order constant contagion}
\end{equation}
where $\rho(t)=\int s_{a,p}^tdadp$.

By rescaling the transmission probability through dividing by $\mu$, i.e., setting $\lambda_1 = \beta_1/\mu$ and $\lambda_2 = \beta_2/\mu$, we can rewrite equation (\ref{2-order constant contagion}) as:
\begin{equation}
\eqalign{
	\frac{d\rho}{dt} = -\rho+\lambda_1a\langle z(z-1)\rangle\rho(1-\rho)+\frac{\lambda_2ap\langle z(z-1)(z-2)\rangle\rho^2(1-\rho)}{2}
}
\label{2-order constant contagion 2}
\end{equation}
Clearly, equation (\ref{2-order constant contagion 2}) has three equilibriums, which are respectively

\begin{equation}
	\eqalign{
		&\rho^*_1 = 0,\\
		&\rho^*_{2\pm} = \frac{\langle k_2\rangle\lambda_2-\langle k_1\rangle\lambda_1\pm\sqrt{(\langle k_1\rangle\lambda_1-\langle k_2\rangle\lambda_2)^2-4\langle k_2\rangle\lambda_2(1-\langle k_1\rangle \lambda_1)}}{2\langle k_2\rangle\lambda_2}
}
\label{theory 3-solution}
\end{equation}
where,
\begin{equation}
	\eqalign{
		&\langle k_1\rangle = a\langle z(z-1)\rangle\\
		&\langle k_2\rangle =\frac{ap\langle z(z-1)(z-2)\rangle}{2}
}
\end{equation}

Furthermore, we analyze the relationship between the stability of equilibriums $\rho^*_1,\rho^*_{2\pm}$ and transmission parameters $\lambda_1,\lambda_2$. Suppose that the initial value of this system $\rho(t=0)=\rho_0>0$.

\begin{itemize}
	\item[(i)]When $\lambda_1>1/\langle k_1\rangle$, $\rho_{2-}^*$ is always negative. So $\rho_{2+}^*$ is the only stable equilibrium point. In these conditions, norms outbreak no matter what $\rho_0$ is.
	\item[(ii)]When $\lambda_1 \le 1/\langle k_1\rangle$ and either $\lambda_1 \le \lambda_c=(2\sqrt{\lambda_2 k_2}-\lambda_2 k_2)/k_1$ or $\lambda_2 \le 1/\langle k_2\rangle$ holds, only $\rho_1^*$ is the stable equilibrium point. Norms go extinct no matter what $\rho_0$ is.
	\item[(iii)]When $\lambda_1 \le 1/\langle k_1\rangle$, $\lambda_1>\lambda_c=(2\sqrt{\lambda_2 k_2}-\lambda_2 k_2)/k_1$ and $\lambda_2>1/\langle k_2\rangle$, $\rho_{1}^*$ and $\rho_{2+}^*$ are all stable equilibrium points. In these conditions, the system is in bistable state where norms outbreak for $\rho_0>\rho_{2-}^*$ while go extinct for $\rho_0<\rho_{2-}^*$.
\end{itemize}

\subsubsection{Heterogeneous situation}
Here, we consider the 2-order contagion on heterogeneous PAD networks. In this situation, activity rate and probability are sampled from a joint probability distribution $H(a,p)$. For this complex situation, we provide the framework of theoretical analysis in \ref{necessary condition}. We find that the norms outbreak if and only if the rescaled transmissibility $\lambda_1$ satisfies the condition:
\begin{equation}
\eqalign{
	\lambda_1=\frac{\beta_1}{\mu}>\lambda_c^{PAD}
}
\end{equation}
where
\begin{equation}\label{he_threshold}
\eqalign{
		&\lambda_c^{PAD}=\frac{2}{\langle z(z-1)\rangle\langle a\rangle+\langle z-1\rangle\sqrt{\Delta}}
}
\end{equation}

\begin{equation}
\eqalign{
		\Delta&=\left[\langle ap\rangle^2+2\langle ap\rangle\langle a(1-p)\rangle+\langle a^2(1-p)^2\rangle \right]\langle z\rangle^2\\
		&+4\left[\langle a^2p(1-p)\rangle-\langle ap\rangle\langle a(1-p)\rangle\right]\langle z\rangle+4(\langle a^2p^2\rangle-\langle ap\rangle^2)
}	
\end{equation}

\subsection{Theoretical and simulation results}\label{sec4}

\subsubsection{Higher-order contagion on homogeneous PAD networks}\label{subsubsec2}

\begin{figure}
	\center
	
	\includegraphics[width=12.5cm]{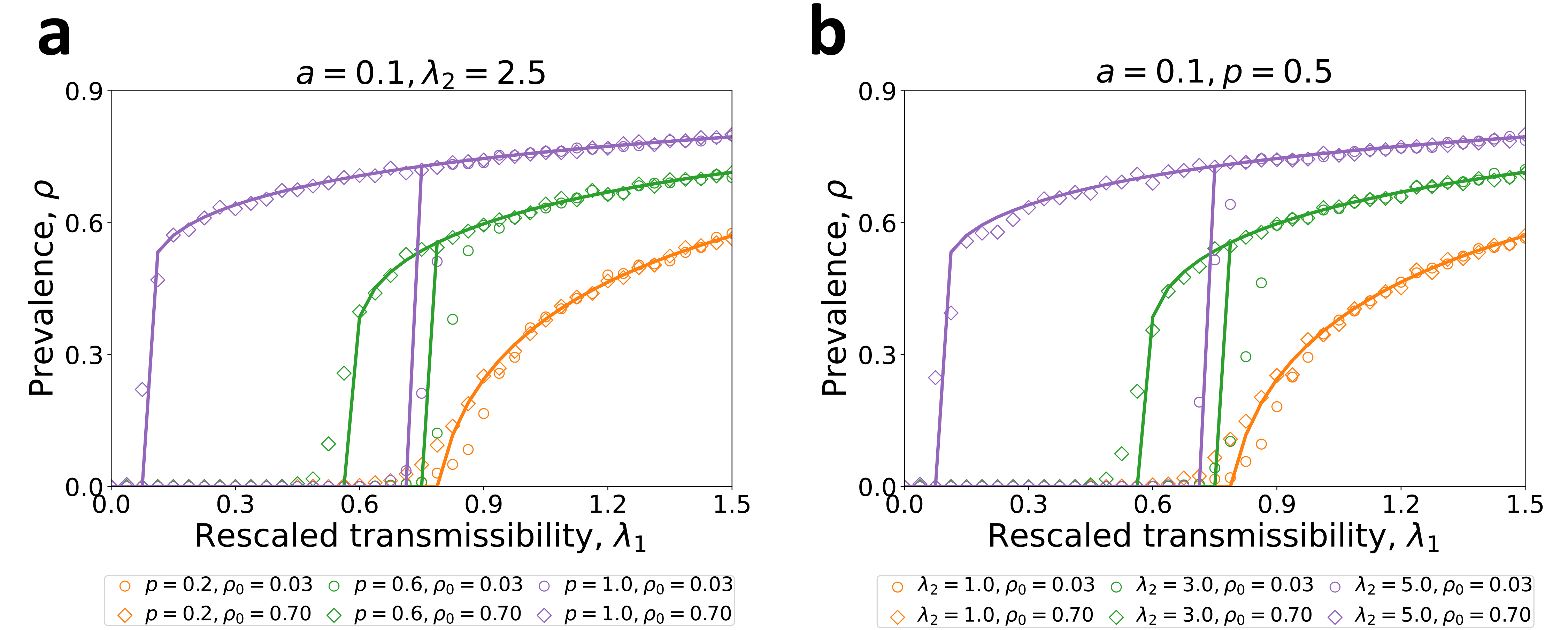}
	
	\caption{Emergence of bistable phenomena. Prevalence curves are shown against rescaled transmissibility $\lambda_1$ (a) under different combinations of $p$ and initial density of spreaders $\rho_0$, and (b)under different combinations of $\lambda_2$ and $\rho_0$. Simulation results are averaged over 20 independent runs. The solid lines correspond to the theoretical predictions solved by equation (\ref{theory 3-solution}). Parameters: (a) $a=0.1, \lambda_2=2.5$; (b) $a=0.1, p=0.5$.}
	\label{sim_constant_s=3}
\end{figure}

In this section, we focus on $2$-order contagion on homogeneous PAD networks, and explore how time-varying characteristics of network structure affect spreading results, especially the emerging  critical phenomena. Consider a PAD network with $N=1500$ nodes, where each node has the same activity rate $a=0.1$ and the possibility of creating higher-order simplex $p$. Considering that $p$ directly determine the higher-order structure of networks, we also called $p$ as higher-order structure parameter. Assume that the order of generated simplex, denoted by $z$, obeys a discrete distribution $p(z=l)=1/3$ $(l=3,4,5)$. Initially, we randomly set a certain fraction of population $\rho_0$ as spreaders.

\begin{figure}
	\centering
	\includegraphics[width=12.5cm]{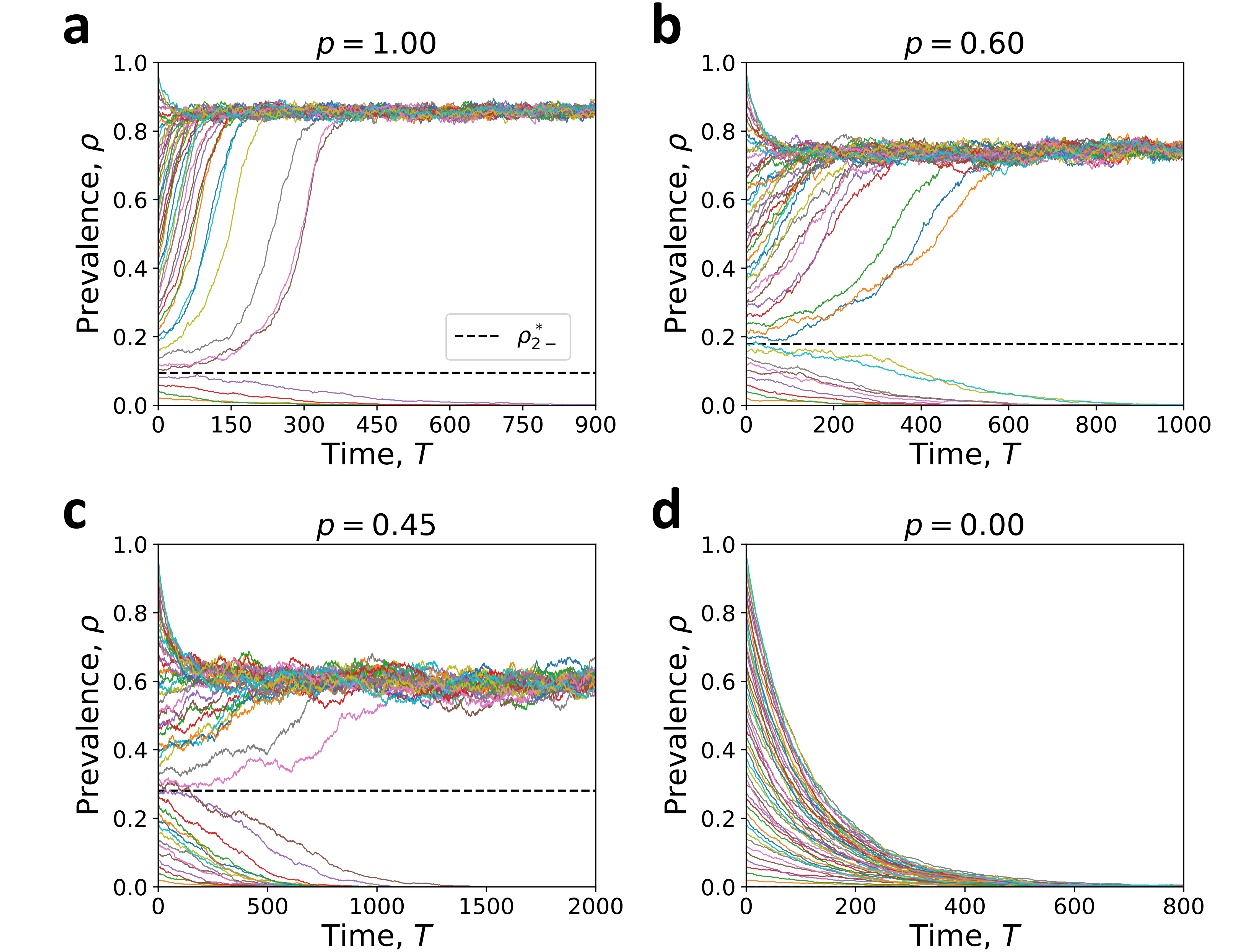}
	
	\caption{Effect of initial density of spreaders. Set the value of higher-order structure parameter $p$ as (a) $p=1$, (b) $p=0.6$, (c) $p=0.45$, (d) $p=0$. Time evolutions of prevalence are shown under different initial density of spreaders $\rho_0$. In each figure, a single curve corresponds to one value of different initial densities of spreaders. The dashed horizontal line corresponds to the unstable branch $\rho_{2-}^*$ solved by equation (\ref{theory 3-solution}), which separates the two final state. Parameters: $a=0.1,\lambda_1=0.2,\lambda_2=6$.}
	
	\label{cnstant_3_time}
\end{figure}

Firstly, we show the prevalence curves as a function of rescaled transmissibility $\lambda_1$ for different combinations of higher-order structure parameter $p$ and initial density of spreaders (figure \ref{sim_constant_s=3}(a)). The case $p=0.2$ (the orange curves) is similar to the SIS model on activity-driven networks, which displays continuous phase transition~\cite{perra2012activity}. However, there emerges bistable phenomena, i.e., two stable equilibriums: outbreak (if $\rho_0=0.70$) and extinction (if $\rho_0= 0.03$), when $p$ is large ($p=0.6, 1$). It indicates that the final prevalence might depend on initial density of spreaders in bistable region. Subsequently, figure \ref{sim_constant_s=3}(b) presents the prevalence against $\lambda_1$ under different higher-order transmission parameter $\lambda_2$. Results show that large $\lambda_2$ also induces the emergence of bistability. Overall, the emergence of bistability might be jointly determined by higher-order transmissibility $\lambda_2$ and higher-order structure parameters $p$. In addition, all subfigures show that our theoretical solutions well predict simulation results.

\begin{figure}
	\centering
	\includegraphics[width=12.5cm]{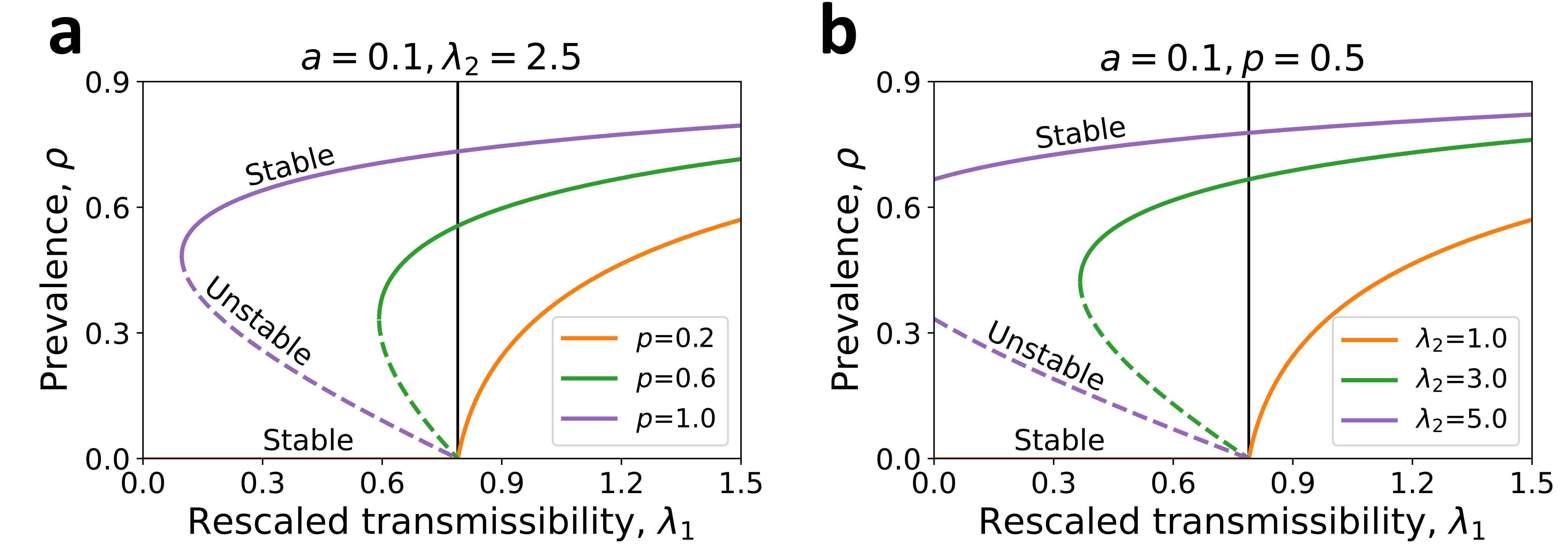}
\caption{Emergence of unstable manifold. Multiple equilibriums given by eq.(\ref{theory 3-solution}) are plotted as a function of rescaled transmissibility $\lambda_1$ under (a) different $p$ and  (b) different $\lambda_2$. Continuous and dashed lines correspond to stable and unstable equilibrium manifolds, respectively. The vertical line represents the threshold $\lambda_1=1/\langle k_1\rangle$. In addition, (a) $a=0.1,\lambda_2=2.5$, (b) $a=0.1,p=0.5$.}
	\label{theory_manifold_s=3}
\end{figure}

In order to further illustrate how initial density of spreaders affects the evolutionary result, we present the time evolutions of prevalence under different initial densities of spreaders (see figure \ref{cnstant_3_time}(a)-(d)). In each subfigure, different curves correspond to different values for the initial density of spreaders. Figure \ref{cnstant_3_time}(a)-(c) present the situation of large $p$. Results show the determinant effect of $\rho_0$ in affecting the final prevalence. Specifically, Our equation (\ref{theory 3-solution}) accurately provides a threshold (dashed line in all figures): prevalence vanishes if $\rho_0$ is smaller than the threshold, while reaches a endemic state if $\rho_0$ is above the threshold. Besides, as probability $p$ decreases, the threshold value for $\rho_0$ increases gradually. Figure. \ref{cnstant_3_time}(d) discusses the situation of $p=0$, i.e., network with no higher-order structure, where the prevalence vanishes no matter what $\rho_0$ is.

\begin{figure}
	\centering
	\includegraphics[width=12.5cm]{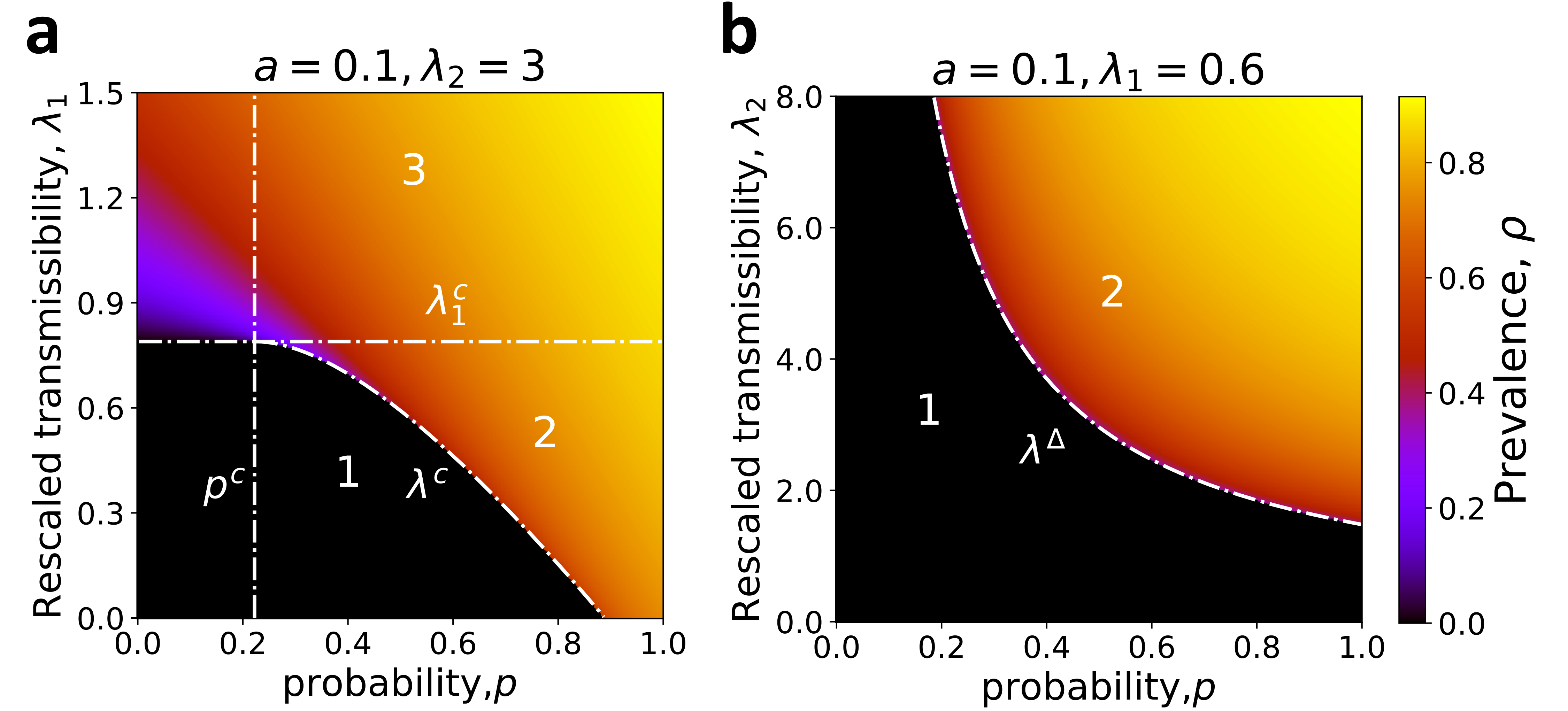}
	\caption{Critical conditions of phase transitions. Shown are phase diagrams about final prevalence under (a) different combinations of $p$ and $\lambda_1$ and under (b) different combinations of $p$ and $\lambda_2$. (a) The phase plane consists of three regions: extinction (region 1), bistable (region 2) and outbreak (region 3). The dash-dotted line $p^c=2/(\lambda_2a\langle z(z-1)(z-2)\rangle),\lambda_1^c=1/\langle k_1\rangle,\lambda^c=(2\sqrt{\lambda_2\langle k_2\rangle }-\lambda_2\langle k_2\rangle)/\langle k_1\rangle$. (b) The phase plane consists of two regions: extinction (region 1), bistable (region 2). The dash-dotted line $\lambda_\Delta=(2+2\sqrt{1-\langle k_1\rangle\lambda_1}-\langle k_1\rangle\lambda_1)/\langle k_2\rangle$. Parameters: (a) $a=0.1$,$\lambda_2=3$; (b) $a=0.1$,$\lambda_1=0.6$.}
	\label{xiangtu_s=3}
\end{figure}

Then, figure \ref{theory_manifold_s=3} shows the changes of all equilibriums against $\lambda_1$, including unstable and stable manifold, under different higher-order structure parameter $p$ and higher-order transmission parameter $\lambda_2$. The stable manifolds (solid lines) again illustrate that large $p$ and $\lambda_2$ can induces the emergence of bistable region. Moreover, results also show the emergence of unstable manifold which appears between two stable manifolds. In particular, the value of unstable equilibriums corresponds to the critical initial density of spreaders, above which norm would outbreak. We find that the prevalence corresponding to unstable equilibriums decreases as $\lambda_2$ or $p$ grows, which is consistent with the simulation results from figure \ref{cnstant_3_time}.

Furthermore, by analyzing the evolutionary equation (\ref{theory 3-solution}), we explore how time-varying characteristics of networks affect the phase transition of the coupling dynamics. Figure \ref{xiangtu_s=3}(a) presents the phase diagram describing the joint effect of $p$ and $\lambda_1$ on the final prevalence. There are three regions corresponding to three phases of evolutionary results: extinction (region 1), bistable (region 2), outbreak (region 3). In particular, the transition from region 1 to region 2 satisfies $\lambda^c=(2\sqrt{\lambda_2\langle k_2\rangle }-\lambda_2\langle k_2\rangle)/\langle k_1\rangle$, where $\langle k_2\rangle$ largely depends on the value of $p$. It indicates that large $p$ is necessary for bistable phenomena. Moreover, we consider how higher-order structure parameter $p$ and higher-order transmission parameter $\lambda_2$ jointly affect the final prevalence. As shown in figure \ref{xiangtu_s=3}(b), we analytically derive the conditions for bistable phenomena, which is $\lambda_2>\lambda_\Delta=(2+2\sqrt{1-\langle k_1\rangle\lambda_1}-\langle k_1\rangle\lambda_1)/\langle k_2\rangle$. It can be concluded that the higher-order contagion dynamics and time-varying characteristics of network jointly determine the appearance of bistable state.

\subsubsection{Higher-order contagion on heterogeneous PAD networks}

In the real-world networks, each person interacts with others in different frequencies and ways, corresponding to the heterogeneous distribution of node activity rate and probability on the PAD model. In this section, we explore complex social contagion on heterogeneous PAD networks. We consider a heterogeneous PAD network with $N=1500$ nodes, which the activity rates sampled from $F(a) \propto a^{-1.7}(a\in\left[0.01,1\right))$ and higher-order structure parameter $p\sim U(0,1)$.

\begin{figure}
	\centering
	\includegraphics[width=12.5cm]{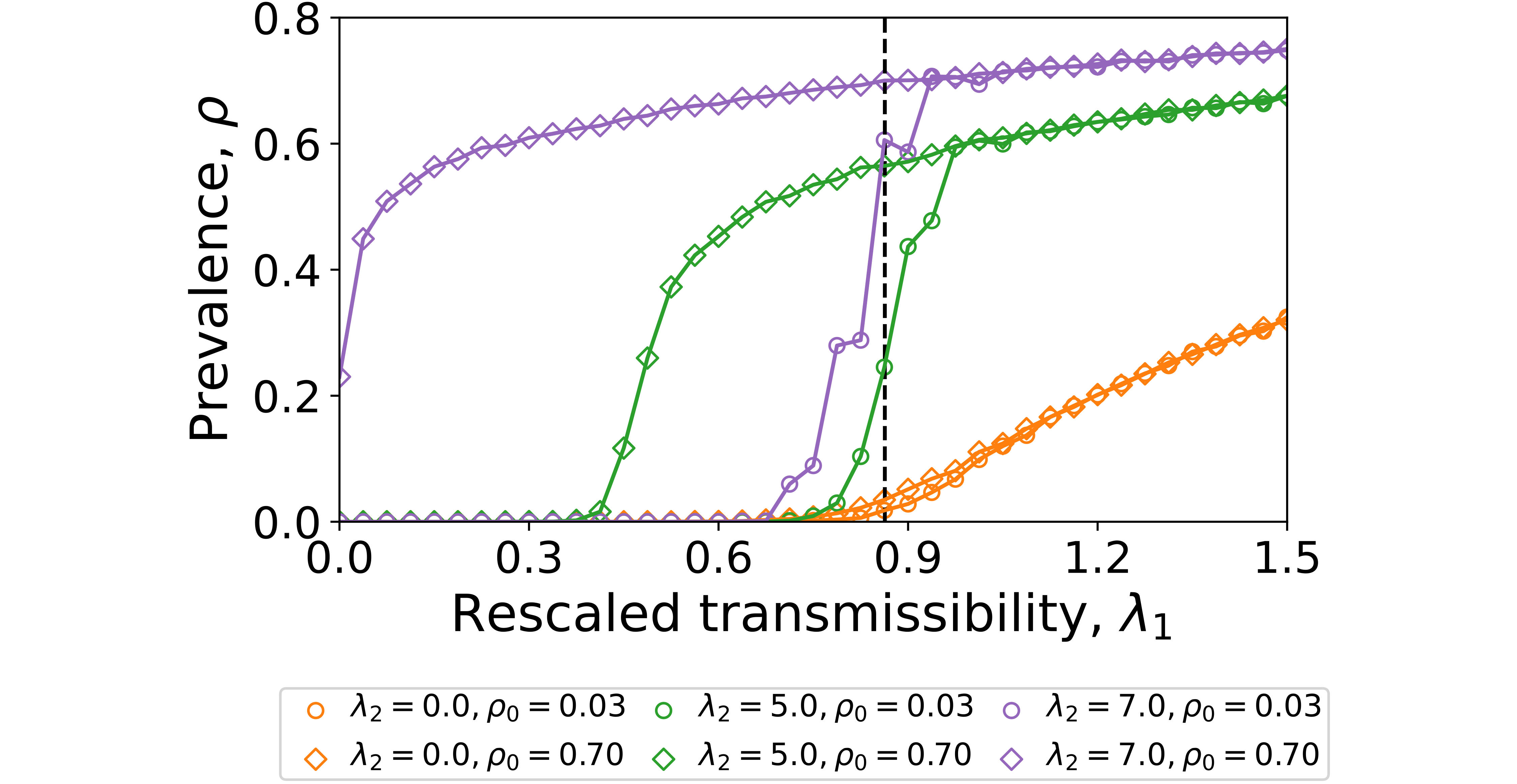}
	\caption{2-order contagion on heterogeneous PAD networks. Prevalence curves are shown against rescaled transmissibility $\lambda_1$ for different values of rescaled transmissibility $\lambda_2$ and initial density of adopters $\rho_0$. The activity rates sampled from $F(a) \propto a^{-1.7}(a\in\left[0.01,1\right))$ and probability $p\sim U(0,1)$. The vertical dotted line corresponds to the threshold described by eq.(\ref{he_threshold}).}
	\label{sim_he_s=3}
\end{figure}

In figure \ref{sim_he_s=3}, the prevalence is presented as a function of rescaled transmissibility $\lambda_1$ for different combinations of rescaled transmissibility $\lambda_2$ and initial density of spreaders $\rho_0$.  Results show the similar phenomena on homogeneous PAD networks, including bistability. In addition, we also note that bistable phenomena will disappear when $\lambda_1>\lambda_c^{PAD}$.

\section{Conclusions and Discussions}

Network modeling aims at generating synthetic networks with specified topological properties, which plays a fundamental role in exploring structure and dynamics of networked systems~\cite{boccaletti2006complex}. Recently, empirical studies show that higher-order dynamics is an indispensable element in social interactions and can significantly affect group structures and functions of complex systems. This leads to new scientific challenges in higher-order network modeling. In particular, it remains unclear how to generate higher-order networks holding both power-law degree distribution and high-clustering characteristics, which are ubiquitous in real-world networks such as various social networks.

To address this problem, we propose probabilistic activity driven model (PAD model) by incorporating the activity rate of nodes and the possibility of creating higher-order interactions (i.e., higher-order structural mechanism). The aggregated networks generated by PAD model have both power-law degree distribution and high-clustering characteristics. Surprisingly, their power-law exponents and clustering coefficients can be tuned in a wide range by altering model parameters associating with individual activity distribution and higher-order structural mechanism. Furthermore, we provide an approximation algorithm to select a group of parameters that can generate networks with any desired structural properties and verify its effectiveness by reconstructing three real-world networks with different topological properties.

In addition, the temporal higher-order networks generated by PAD model can co-evolve with higher-order dynamics at the same time scale. We investigate how the network evolution affects higher-order contagion processes, especially the shifts of phase transitions. We find the emergence of bistable phenomena where outbreak and extinction coexist, which is determined by the joint effect of higher-order structural mechanism and higher-order transmission dynamics. Moreover, we provide a theoretical framework describing such coupling dynamics and analytically derive the critical conditions.

PAD model provides a simple way to generate networks approximating the given power-law exponents and clustering coefficient, which has great potential in exploring higher-order dynamics on real-world networks. Our work also provides important insights toward how temporal higher-order structure affects contagion dynamics. Other dynamical processes such as synchronization~\cite{kohar2014synchronization} and evolutionary games~\cite{alvarez2021evolutionary} on the top of temporal PAD networks, as well as other more complicated mechanisms in network generating processes such as memory effects~\cite{zino2018modeling} and node attractiveness~\cite{pozzana2017epidemic}, are worthy of further consideration.     

\ack{This work is supported by Program of National Natural Science Foundation of China Grant No. 12201026, 11871004, 11922102, 62141605 and National Key Research and Development Program of China Grant No. 2018AAA0101100, 2021YFB2700304.}


\providecommand{\newblock}{}

\newpage
\pagestyle{empty}

\begin{appendix}

\section{Derivation of degree distribution of aggregated network}\label{distribution}

\setcounter{equation}{0}
\renewcommand{\theequation}{A.\arabic{equation}}

In an aggregated network with the aggregation steps $T$, the probability of connection between node $i$ and node $j$ consists of three parts:

Node $i$ is actively connected to node $j$ for event $A$, where $\langle m\rangle=\langle z(z-1)\rangle/2$.

\begin{equation}
	P(A)=1-\left[1-\frac{\langle z-1\rangle p_i+(1-p_i)\langle m\rangle}{N-1}\right]^{a_iT}
\end{equation}

Node $j$ is actively connected to node $i$ for event $B$,

\begin{equation}
	P(B)=1-\left[1-\frac{\langle z-1\rangle p_j+(1-p_j)\langle m\rangle}{N-1}\right]^{a_jT}
\end{equation}

Node $k$ is actively connected to both node $i$ and node $j$ for event $C$,

\begin{equation}
	\eqalign{
		P(C)&=1-(1-\frac{\langle (z-1)(z-2)\rangle}{(N-1)(N-2)})^{\sum_{k\not=i,j}a_kp_kT} \\
		& \simeq 1-\left[1-\frac{\langle (z-1)(z-2)\rangle}{(N-1)(N-2)}\right]^{(N-2)\langle ap\rangle T}
}
\end{equation}

Therefore, the probability of connection between node $i$ and node $j$ is,
\begin{equation}
	\eqalign{
		&P(A\cup B\cup C)=1-\left[1-\frac{\langle z-1\rangle p_i+(1-p_i)\langle m\rangle}{N-1}\right]^{a_iT}\\
		&\left[1-\frac{\langle z-1\rangle p_j+(1-p_j)\langle m\rangle}{N-1}\right]^{a_jT}\left[1-\frac{\langle (z-1)(z-2)\rangle}{(N-1)(N-2)}\right]^{(N-2)\langle ap\rangle T}
	}
\end{equation}

In the mean-field approximation, the degree of node $i$ in the aggregated PAD network is,

\begin{equation}\label{appendix degree equation}
	\eqalign{
		&k_i^{PSAD}(T)=(N-1)\{1-\left[1-\frac{\langle z-1\rangle p_i+(1-p_i)\langle m\rangle}{N-1}\right]^{a_iT}\\
		&\left[1-\frac{\langle z-1\rangle\langle p\rangle+\langle1-p\rangle \langle m\rangle}{N-1}\right]^{\langle a\rangle T}\left[1-\frac{\langle (z-1)(z-2)\rangle}{(N-1)(N-2)}\right]^{(N-2)\langle ap\rangle T}\}\\
		&\simeq N(1-e^{-\frac{T(a_id_i+\langle a\rangle\langle d\rangle+h)}{N}})
	}
\end{equation}

where the approximation holds for $N\gg 1$, $d_i=\langle z-1\rangle p_i+(1-p_i)\langle m\rangle$, $\langle d\rangle=\langle z-1\rangle\langle p\rangle+\langle 1-p\rangle\langle m\rangle$, $h=\langle (z-1)(z-2)\rangle\langle ap\rangle$.

Then we assume that all nodes have the same $p$. Using the expressions above, we obtain the functional form for the degree distribution:
\begin{equation}\label{or 1-order degree distribution}
	P_T(k)\sim\frac{1}{T\langle d\rangle(1-\frac{k}{N})}F\left[-\frac{N}{T\langle d\rangle}\ln(1-\frac{k}{N})-\langle a\rangle-\frac{h}{\langle d\rangle}\right]
\end{equation}

Considering that $N\gg k$ in real networks, equation (\ref{or 1-order degree distribution}) can be simplified as:
\begin{equation}
	P_T(k)\sim\frac{1}{T\langle d\rangle}F\left[\frac{k}{T\langle d\rangle}-\langle a\rangle-\frac{h}{\langle d\rangle}\right]
\end{equation}

\begin{algorithm}[]\label{fit-algorithm}
	\caption{Real-world Network Fitting}
	\label{alg:Framwork}
	\begin{algorithmic}[1]
		\renewcommand{\algorithmicrequire}{ \textbf{Input:}} 
		\renewcommand{\algorithmicensure}{ \textbf{Output:}}
		\Require  
		Real-world Network $G_{real}$; Error $\epsilon$
		\renewcommand{\algorithmicrequire}{ \textbf{Function:}}
		\Ensure 
		Fitting Network $G_{fit}$
		\Require
		Computational structural properties, $NetAnalyse$; Initialization parameters, $Init$; Parameter adjustment, $Adjust$; Aggregated network, $AGG$; Largest connected component, $GCC$; PAD network generation, $PAD$; Signum function, $Sgn$
		\Function{$Adjust$}{$u,\epsilon$}
		\If {$(u['N'])^2>0.25\epsilon$}
		\State $N\mathrel{+}=(Sgn(u['N'])*\Delta N)$
		\EndIf
		\If {$(u['\langle k\rangle'])^2>0.25\epsilon$}
		\State $T\mathrel{+}=(Sgn(u['T'])*\Delta T)$
		\EndIf
		\If {$(u['\gamma'])^2>0.25\epsilon$}
		\State $\eta\mathrel{+}=(Sgn(u['\gamma'])*\Delta \eta)$
		\EndIf
		\If {$(u['\langle c\rangle'])^2>0.25\epsilon$}
		\State $p\mathrel{+}=(Sgn(u['\langle c\rangle'])*\Delta p)$
		\EndIf
		\If {$p>1$}
		\State \Return $\textbf{False}$
		\Else
		\State \Return $\textbf{True}$
		\EndIf
		\EndFunction
		\State $SP_{real}=NetAnalyse(G_{real})$;
		\State $\eta,\epsilon,N,p,z,P(z)=Init()$;
		\State $T=\lfloor\frac{SP_{real}['\langle k\rangle']}{2\langle a\rangle\langle m\rangle}\rfloor$, $u=None$;
		\While{($\textbf{not}\ u$) $\textbf{or}$ ($\langle u.values(),u.values()\rangle>\epsilon$)}
		\If{$u$}
		\If {$\textbf{not}\ Adjust(u,\epsilon)$}
		\State \Return $\textbf{False}$
		\EndIf
		\EndIf
		\State $G_T=AGG[PAD(\eta,\epsilon,N,p,z,P(z),T)]$
		\State $G_{fit}=GCC(G_T)$
		\State $SP_{fit}=NetAnalyse(G_{fit})$
		\State $u=\frac{SP_{real}-SP_{fit}}{SP_{real}}$
		\EndWhile \\
		\Return $G_{fit}$; 
	\end{algorithmic}
\end{algorithm}

\section{Real-world network fitting method}\label{fitting method}

In the algorithm above, the input is the real-world network $G_{real}$ and the allowed error $\epsilon$, and the output is the fitted network $G_{fit}$. Firstly, we calculate the structural properties of $G_{real}$. Next, we initialize the parameters according to figure $\ref{phase structure p k}$ and $SP_{real}$. When $T$ is small, the average degree $\langle k\rangle$ of the aggregated PAD network can be approximated as $2\langle a\rangle\langle m\rangle T$, so we initialize $T$ as $\frac{SP_{real}['\langle k\rangle']}{2\langle a\rangle\langle m\rangle}$. Then, based on parameters, the PAD model generate $G_{fit}$. Finally, we analyze the difference between $SP_{real}$ and $SP_{fit}$ to adjust the parameters, and output the fitting network $G_{fit}$ when the error is lower than $\epsilon$.

\section{Real-world network data description}\label{data descripution}

\setcounter{equation}{0}
\renewcommand{\theequation}{C.\arabic{equation}}

$Deezer\enspace Europe\enspace Social\enspace Network\enspace$ A social network of Deezer users which was collected from the public API in March 2020. Nodes are Deezer users from European countries and edges are mutual follower relationships between them.

$Deezer\enspace Croatia\enspace Social\enspace Network\enspace$ The data was collected from the music streaming service Deezer (November 2017). The dataset represents friendship networks of users from Romania. Nodes represent the users and edges are the mutual friendships.

$LastFM\enspace Asia\enspace Social\enspace Network\enspace$ A social network of LastFM users which was collected from the public API in March 2020. Nodes are LastFM users from Asian countries and edges are mutual follower relationships between them.

\section{Necessary condition for bistability on heterogeneous PAD networks}\label{necessary condition}

\setcounter{equation}{0}
\renewcommand{\theequation}{D.\arabic{equation}}

Inserting $k=2$ in equation (\ref{social contagion equation}), we obtain the equation,

\begin{equation}\label{heterogeneous 2-order equation}
	\eqalign{
		&s_{a,p}^{t+\Delta t}-s_{a,p}^t = -\mu \Delta ts_{a,p}^t + \beta_1\Delta ti_{a,p}^ta(1-p)\int P(m)m dm\int s_{a^{'},p^{'}}^tda^{'}dp^{'}\\
		&+\beta_1\Delta ti_{a,p}^t\int P(m)m dm\int a^{'}(1-p^{'}) s_{a^{'},p^{'}}^tda^{'}dp^{'}\\
		&+\sum_{w=1}^{2} \beta_w \Delta t i_{a,p}^tap\int P(z){z-1\choose w}dz(\int s_{a^{'},p^{'}}^tda^{'}dp^{'})^w\\
		&+\sum_{w=1}^{2} \beta_w\Delta t i_{a,p}^t\int a^{'}p^{'}s_{a^{'},p^{'}}^tda^{'}dp^{'}\int P(z)(z-1){z-2\choose w-1}dz(\int s_{a^{''},p^{''}}^tda^{''}dp^{''})^{w-1}\\
		&+\sum_{w=1}^{2} \beta_w \Delta ti_{a,p}^t\int a^{'}p^{'}n_{a^{'},p^{'}}^tda^{'}dp^{'}\int P(z)(z-1){z-2\choose w}dz(\int s_{a^{''},p^{''}}^tda^{''}dp^{''})^{w}
	}
\end{equation}

Then, we explore the necessary conditions for bistability. When the initial density of spreaders is very small, the higher-order terms can be ignored.

We integrate equation (\ref{heterogeneous 2-order equation}) over $a$ and $p$, then we get an equation for $\rho^t$ that ignore the higher-order terms:

\begin{equation}
	\eqalign{
		&\rho^{t+\Delta t}-\rho^t=-\mu \Delta t\rho^t+\beta_1\Delta t\langle m\rangle \langle a(1-p)\rangle \rho^t+\beta_1\Delta t\langle m\rangle\phi^t+\\
		&\beta_1\Delta t\langle z-1\rangle\langle ap\rangle \rho^t+\beta_1\Delta t\langle z-1\rangle\theta^t+\beta_1\Delta t\langle (z-1)(z-2)\rangle\langle ap\rangle \rho^t
	}
\end{equation}

where $\theta^t=\int aps_{a,p}^tdadp$, $\phi^t=\int a(1-p)s_{a,p}^tdadp$.

Multiplying equation (\ref{heterogeneous 2-order equation}) by $ap$ and integrating we get an equation for $\theta^t$ that ignore the higher-order terms:
\begin{equation}
	\eqalign{
		\theta^{t+\Delta t}-\theta^t&=-\mu \Delta t\theta^t+\beta_1\Delta t\langle m\rangle\langle a^2p(1-p)\rangle \rho^t+\beta_1\Delta t\langle m\rangle\langle ap\rangle\phi^t\\
		&+\beta_1\Delta t\langle z-1\rangle\langle a^2p^2\rangle \rho^t+\beta_1\Delta t\langle z-1\rangle\langle ap\rangle\theta^t\\
		&+\beta_1\Delta t\langle (z-1)(z-2)\rangle\langle ap\rangle\rho^t
	}
\end{equation}

Multiplying equation (\ref{heterogeneous 2-order equation}) by $a(1-p)$ and integrating we get an equation for $\phi^t$ that ignore the higher-order terms:
\begin{equation}
	\eqalign{
		&\phi^{t+\Delta t}-\phi^t=-\mu\Delta t\phi^t+\beta_1\Delta t\langle m\rangle\langle a^2(1-p)^2\rangle \rho^t+\beta_1\Delta t\langle m\rangle\langle a(1-p)\rangle\phi^t\\
		&+\beta_1\Delta t\langle z-1\rangle\langle a^2p(1-p)\rangle \rho^t+\beta_1\Delta t\langle z-1\rangle\langle a(1-p)\rangle\theta^t\\
		&+\beta_1\Delta t\langle (z-1)(z-2)\rangle\langle ap\rangle\langle a(1-p)\rangle \rho^t
}
\end{equation}

These equations can be rewritten as
\begin{equation}
	\left(\begin{array}{c}
		\rho^{t+1}-\rho^t \\
		\theta^{t+1}-\theta^t \\
		\phi^{t+1}-\phi^t
	\end{array}
\right)
	=J
	\left(\begin{array}{c}
		\rho^t \\
		\theta^t \\
		\phi^t
	\end{array}
\right)
\end{equation}

with

\begin{equation}
	J=
	\left(\begin{array}{ccc}
		A & \beta_1\langle z-1\rangle & \beta_1\langle m\rangle\\
		B & -\mu+\beta_1\langle z-1\rangle\langle ap\rangle & \beta_1\langle m\rangle\langle ap\rangle\\
		C & \beta_1\langle z-1\rangle\langle a(1-p)\rangle & -\mu+\beta_1\langle m\rangle\langle a(1-p)\rangle
	\end{array}
\right)
\end{equation}

where,

\begin{equation}
	\eqalign{
		&A=-\mu+\beta_1\langle m\rangle\langle a(1-p)\rangle+\beta_1\langle z-1\rangle^2\langle ap\rangle\\
		&B=\beta_1\langle m\rangle\langle a^2p(1-p)\rangle+\beta_1\langle z-1\rangle\langle a^2p^2\rangle+\beta_1\langle (z-1)(z-2)\rangle\langle ap\rangle^2\\
		&C=\beta_1\langle m\rangle\langle a^2(1-p)^2\rangle+\beta_1\langle z-1\rangle\langle a^2p(1-p)\rangle\\
		&+\beta_1\langle (z-1)(z-2)\rangle\langle ap\rangle \langle a(1-p)\rangle
	}
\end{equation}

its eigenvalues,
\begin{equation}
	\eqalign{
		&\kappa_0=-\mu \\
		&\kappa_{\pm}=\frac{\beta_1\langle z(z-1)\rangle(\langle ap\rangle+\langle a(1-p)\rangle)-2\mu}{2}\pm\langle z-1\rangle\beta_1\frac{\sqrt{\Delta}}{2}
	}
\end{equation}

where,
\begin{equation}
	\eqalign{
		\Delta=&\left[\langle ap\rangle^2+2\langle ap\rangle\langle a(1-p)\rangle+\langle a^2(1-p)^2\rangle \right]\langle z\rangle^2\\
		&+4\left[\langle a^2p(1-p)\rangle-\langle ap\rangle\langle a(1-p)\rangle\right]\langle z\rangle+4(\langle a^2p^2\rangle-\langle ap\rangle^2)
}	
\end{equation}

\pagestyle{empty}

The norms outbreak if and only if the largest eigenvalue $\kappa_+$ is positive. This yields the rescaled transmissibility $\lambda_1$ threshold condition:
\begin{equation}
	\frac{\beta_1}{\mu}>\lambda_c^{PAD}
\end{equation}

with:

\begin{equation}\label{he_yuzhi}
	\eqalign{
		&\lambda_c^{PAD}=\frac{2}{\langle z(z-1)\rangle\langle a\rangle+\langle z-1\rangle\sqrt{\Delta}}
	}
\end{equation}

\end{appendix}

\end{document}